# Artificial light at night: a global disruptor of the nighttime environment

Salvador Bará[1,*] and Fabio Falchi[1,2]

[1] *Departamento de Física Aplicada, Universidade de Santiago de Compostela (USC), Santiago de Compostela, 15782 Galicia (Spain)*
[2] *ISTIL Istituto di Scienza e Tecnologia dell'Inquinamento Luminoso –Light Pollution Science and Technology Institute, I-36016 Thiene, Italy*

## Summary

Light pollution is the alteration of the natural levels of darkness by an increased concentration of light particles in the nighttime environment, resulting from human activity. Light pollution is changing in a deep way the environmental conditions of the night in wide areas of the planet, and is a relevant stressor whose effects on life are being unveiled by a compelling body of research. In this paper we briefly review the basic aspects of artificial light at night as a pollutant, describing its character, magnitude and extent, its worldwide distribution, its temporal and spectral change trends, as well as its dependence on current light production technologies and prevailing social uses of light. It is shown that the overall effects of light pollution are not restricted to local disturbances, but give rise to a global, multiscale disruption of the nighttime environment.

## 1. Introduction

Artificial light is a key resource for humankind. It provides our societies with an unprecedented freedom from the natural illumination cycles, enabling the extension of human activities into nighttime and the use during the day of many indoor spaces lacking enough natural light. A long and rich history of technological breakthroughs enabled the production and control of ever-increasing amounts of light since ancient times [1], a process intensely accelerated in the last centuries [2, 3]. The industrial supply of artificial light is nowadays an integral feature of many societies, and its widespread availability is so taken for granted that lighting has become part of the 'invisible infrastructures' [4, 5], essential for everyday life but whose existence is properly noticed only if (and when) they fail.

The massive use of this resource, however, comes at a price. Oudoor light sources of any type, as well as indoor sources whose light spills outdoors, increase the concentration of light particles (photons) in the environment [6], giving rise to light pollution. Light directly propagated from the lamps, reflected on surfaces, or scattered in turbid media (air and water) contribute to this increase. Photons know no borders [7] and a significant fraction of them travel long distances [8-11] before interacting with atmospheric constituents, inert surfaces, or living matter. As a result, many areas surrounding cities do not experience the true darkness of the night but face a kind of permanent twilight. The *New world atlas of artificial night sky brightness* shows that more than 80% of the world population live under light-polluted skies and 23% of the world's land surfaces between 75°N and 60°S experience light-polluted nights [12]. Recent estimates of the increasing artificial light emissions and the decreasing number of stars visible with the unaided eye (see section 3) strongly suggest that the situation may have deteriorated since the New World Atlas was published (2016). The loss of the natural night is not restricted to terrestrial areas and their aerial space [13]; it is also experienced in the seas [14-17] and freshwaters [18-20], at their surfaces [12, 21-23], air space [24, 25] and down along the water column [26-30].

*Author for correspondence (salva.bara@usc.gal). Former profesor titular at USC (retired).
SB ORCID: 0000-0003-1274-8043, FF ORCID: 0000-0002-3706-5639

Light-matter interactions mediate a wealth of physiological and behavioral processes, including metabolism, foraging, reproduction, predation, and migrations [31]. Light provides energy for sustaining basic processes in animals and plants, and carries information both visual (space awareness) and non-visual (circadian and seasonal entrainment) [32, 33]. Expectedly, life in the photic sphere of the Earth developed efficient adaptations to anticipate and take advantage of the periodic cycles of light and darkness. It is not surprising that the introduction of artificial light in the environment disrupts in meaningful ways the ecological nightscape in which life evolved and thrived during millions of years [34]. Artificial light effects take place at multiple size and complexity scales, ranging from molecular to ecosystem level [31].

Fragmentary examples of phototactic responses elicited by artificial light were already described in ancient literatures [35], and detailed accounts of the disruptive effects of artificial light on some taxa were reported in the 19th century [36]. Substantial advances in the understanding of light pollution as an environmental problem were made in the second half of the 20th century [37-40]. This approach gained widespread recognition in 2004, when Travis Longcore and Catherine Rich coined the concept "ecological light pollution" [41] paving the way for a wider and better appraisal of the polluting character of artificial light. Nowadays, environmental light pollution studies are a thriving field of research [32-34, 42-48] whose findings add to a growing body of knowledge on the disruptive effects of artificial light at night in ground-based astronomy [49, 50], cultural heritage [51, 52], and public health [53-56]. A comprehensive review of the environmental light pollution literature since its modern onset up to present times would clearly demand more space than is available in this paper [57, 58]. An updated list of references can be found in the *Artificial light at night literature database* [59].

In this paper we formalize some aspects of artificial light as a 'proper' pollutant (section 2). We provide a general overview of its spatial and temporal distribution (section 3), its spectral change trends (section 4), as well as its dependence on current light production technologies and prevailing social uses of light (section 5). Additional remarks and conclusions are drawn in section 6.

## 2. A pollutant hidden in plain sight

The term "light pollution" became commonplace in the last third of the 20th century, when serious concerns on the negative effects of artificial light at night for astrophysical research were expressed by the astronomy community [60-62]. However, during a relatively long period of time the word "pollution" conveyed in this context a sense of relevant nuisance, rather than the action and effect of a pollutant proper. Light pollution was deemed substantially different from 'hard' contamination types, like the chemical pollution of air or water [61]. This misconception left its footprint on several working definitions of light pollution, which frequently contained a somewhat confusing mix of causes and effects, descriptive lists of particular manifestations of the problem (glare, skyglow, light trespass, clutter...), and wrong dichotomies (useful light vs. polluting light). Some examples of this kind are, e.g., "artificial light that shines where it is neither wanted, nor needed" [63] or "the inappropriate or excessive use of artificial light" [64]. Other definitions are analyzed in [65]. However, the polluting character of artificial light *per se* can no longer be ignored. In this section we describe how artificial light at night fulfils the conditions to be considered a classical, standard pollutant.

### (a) Artificial light at night is an actual pollutant
Leaving aside narrow or ad-hoc definitions intended for specific applications, usual scientific and legal practice consider pollutants those

        (i) forms of matter or energy that
        (ii) are produced by human activity,
        (iii) are present in the environment in concentrations that alter the natural ones, and
        (iv) cause or may reasonably cause harm to humans, the environment, and/or other goods of any nature,

see, e.g. [66, 67, 6] for their realization in the UN Convention on Long-range Transboundary Air Pollution (1979). According to the available body of knowledge (see Introduction above) it would be difficult to argue that artificial light at night does not fulfill conditions (i) to (iv). In fact it does, in the same sense as they are met by conventional (e.g. chemical) pollutants.

Every pollutant has specific characteristics, and artificial light at night is not an exception. Artificial light is a mid-range pollutant, capable of producing measurable environmental effects at multiple distance scales, from the immediate neighbourhood of the lamps up to hundreds of km away. It is composed of particles (photons) that carry the smallest possible bits of electromagnetic energy [68]. Photons propagate at the speed $c=299\ 792\ 458$ m·s$^{-1}$ in vacuum. The average time spent by a photon in the terrestrial atmosphere is short, typically ranging from microseconds to milliseconds before they interact with atmospheric constituents, terrain, built environment, any form of living matter, or leave for outer space. Notwithstanding these short effective lives, the concentration of artificial photons in the atmosphere is often stationary or slowly varying in time, due to the huge amounts of light particles continuously produced by artificial sources (see section (b) below). The physical composition of the natural atmosphere becomes that way modified at nighttime. Although the photon number concentration rapidly vanishes if the light sources are switched off, the disruptive effects already produced in the environment may take much longer to disappear (e.g. an area might not be immediately recovered if light-induced habitat fragmentation gave rise to biodiversity loss). It is a well known fact that light gives rise to different environmental effects depending on its wavelength, direction of propagation and polarization. This allows to define appropriate photon classes for the problems under study, binned according to their momentum-energy (wavelength and direction of propagation) and polarization modes.

## (b) Expressing light pollution in terms of the concentration of artificial light particles

The usual way of specifying illumination has somehow contributed to obscure its deep connection with other types of pollutants, whose presence in the environment is commonly expressed in terms of concentrations (e.g. number of molecules per unit volume). Lighting levels, however, can also be described in a natural way in terms of concentrations.

Light is commonly measured in terms of energy or photon flux densities per unit area around a point (irradiances, in W·m$^{-2}$ or photon·s$^{-1}$·m$^{-2}$), per unit solid angle around a propagation direction (radiant intensities, in W·sr$^{-1}$ or photon·s$^{-1}$·sr$^{-1}$), or both (radiances, in W·m$^{-2}$·sr$^{-1}$ or photon·s$^{-1}$·m$^{-2}$·sr$^{-1}$), integrating their spectral distributions (densities per unit wavelength interval) [69] within the relevant spectral passbands. In lighting engineering and for some applications in visual science these radiant quantitites tend to be specified in human-centric photometric units (lx, cd, and cd·m$^{-2}$, respectively) based on the SI unit for luminous intensity, the candela [70]. Detailed descriptions of radiometric and photometric quantities pertinent for ecological studies can be found in [17, 69, 71].

As shown in [6], every radiant quantity defined in terms of photon flux densities can be equivalently expressed in terms of photon concentrations per unit volume. For instance, one lux (lx) is equivalent to $1.36×10^7$ photons per cubic meter propagating towards the measurement surface, within the human visual band. The photon concentrations for other radiometric quantities and spectral bands can be easily calculated. The equivalence of flux densities and concentrations is complete and both descriptions are equally correct, valid and interchangeable. This means that we can keep the customary units for specifying light in environmental and visual science (no change in usual practice is required), and at the same time keep in mind that we are effectively dealing with increased concentrations of particles in the atmospheric or aquatic environment, like we do with other types of pollutants.

## (c) 'All' artificial light at night is pollutant or wasted

On the other hand, the wrong dichotomy of useful versus polluting light, present in some light pollution definitions and coded in lighting regulations [72], overlooks the basic fact that practically all light emitted by outdoor lighting installations, even 'perfect' ones, is pollutant, if not wasted by absorption in urban surfaces. Lighting engineering criteria consider 'useful light' that which illuminates the areas intended to be illuminated, at the right times and with the right intensity and spectrum. A perfect lighting installation, from an engineering viewpoint, is the one that only produces useful light. Adhering to good lighting practices to approximate as much as possible actual installations to 'perfect' ones according to this definition is of course an appropriate and recommended practice. However, let us consider what happens to the light once it reaches the areas intended to be illuminated. First, a big share of the incident photons (~80%) is directly absorbed by these surfaces, whose reflectance in the optical band tends to be within the range 10%-30% for common pavement and façade materials (Fig 1a). Second, and more important from a light pollution viewpoint, only an exceedingly tiny fraction of the reflected photons are actually useful for human vision. As shown in [73], less than 1 photon out of every 22 million reflected ones is captured by the eye of an average observer at 10 m from the reflection point (Fig 1b). The remaining ones, that is, the practical totality in environmental terms, propagate in other directions, being finally absorbed or ending in places that were not intended to be lit, thus contributing to the build up of the total light pollution affecting these sites. This "1/22M+" is an essential limit that cannot be overcome by technological improvements applicable to urban lighting at large scale. In practical terms, then, 'all' light used outdoors is polluting, when not directly lost.

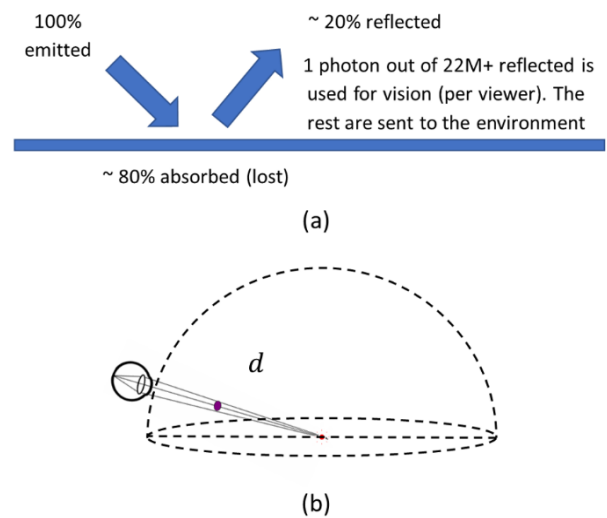

Fig 1. (a) A ~20% of artificial light incident on urban surfaces is reflected, but only an extremely small fraction of it is actually used for human vision, the rest is polluting or lost; (b) The eye only captures those reflected photons propagating in the cone of directions defined by the reflection point and the eye pupil rim. For isotropic reflecting materials this is a fraction $\eta = r^2/(2d^2)$ of the total number of reflected photons, being $r$ the eye pupil radius and $d$ the distance to the point [73].

## (d) Polluting agents and polluted media

It seems advisable to establish an operative distinction between polluting agents and the media that are polluted resulting from their presence. Chemical or biological contaminants, particulate matter, sound and light, among others, are pollutants that affect the quality of the atmospheric and underwater environments. Within this framework, artificial light at night should legitimately be considered an air and water pollutant.

## 3. Spatial distribution and temporal dynamics

The spatial distribution of artificial light sources is strongly correlated with urbanized spaces, connective roadways, infrastructures (factories, ports and airports, …) and any other areas where human activity is assumed or expected at nighttime. This enables using remote sensing of nighttime lights as a proxy indicator for urbanization [74] and economic activity [75]. The widespread mesh of artificial light sources directly pollutes its immediate surroundings, altering substantially their natural darkness and being a relevant factor for habitat fragmentation via phototactic effects that disrupt nighttime ecological corridors [76-78].

The spatial distribution of their polluting effects, however, is not limited to short distances. Artificial light may reach very distant places propagating through the atmosphere. Following basic physical principles of light propagation the resulting light distributions are smoothed by atmospheric scattering and by the superposition of the effects of myriads of individual sources. The light pollution maps can then be though of as "blurred" versions of the sources' maps. This is apparent at the macro-scale by comparing satellite nighttime images showing the sharply defined structure of sources [79-81] with the much more continuous and smoother maps corresponding to their polluting effects, as e.g. the ones shown in [12, 82] for several artificial night sky brightness indicators. The extension of human settlements, the increase in emissions and the broadening effects of the atmosphere cooperate to the progressive encroachment of natural protected spaces [15, 83] and to the deterioration of former pristine sites in which reference astronomical observatories are located [49, 50].

The temporal dynamics of light pollution shows a characteristic multiscale behavior. Firstly, there is a long-term trend of increasing emissions that, depending on the world region and estimation methodology, spans the range 0%-20% per year [43], with independent estimates suggesting an average increase of radiance of order 1.8%-2.2% per year during 2012-2016 in the panchromatic VIIRS-DNB band (500-900 nm) [3]. This is consistent with recent results from the period 1992 to 2017, pointing to an increase of the panchromatic band emissions by at least 49% during these 25 years, that is, about a 1.6% annual, with an estimated increase of the radiance in the visible spectrum of order 270%-400% during the same period on specific places [84]. Note that the rate of deterioration of the night sky visually detected by humans may be significantly faster than the rate of increase of emissions measured by satellites. The last available results in the human visual band suggest actual increases in the range 7%-10% per year during the period 2011-2022, with a world average of order 9.6% [85]. This different progression rates, that could also be expected for other ecologically relevant photometric bands, may be explained by a combination of several factors including changes in the spectral composition of the sources [84, 86], see section 4, and changes in the source number, type or structure that give rise to effects detectable from ground but not from the nadir view of the satellites [85, 87, 88], as e.g. the accelerated installation of ultra-bright LED billboards in central urban areas.

Secondly, there is a seasonal trend in light emissions associated with changes in human activity and ground albedo (related to snow or vegetation cover) [89, 90]. Thirdly, weekly patterns dependent on the workdays cycle as well as hourly patterns along the night are also easily discernable [91, 92]. Emergency situations and catastrophic events frequently give rise to sudden changes in emissions [93-95]. Emissions may also change after urban lighting remodellings driven by energy savings goals [96, 97]. Simultaneous to these human-driven changes in emissions, the highly variable state of the atmosphere (aerosol concentration and types, presence of clouds, ...) [98-104], with different characteristic time scales, determines how the light polluting photons will propagate to long distances from the sources until reaching their final destinations.

## 4. Nightscape blues: the shifting color of the artificial night

One of the most noticeable features of the global outdoor lighting transition experienced in the last two decades is the spectral shift of the emitted light towards shorter (bluer) wavelengths. The yellowish and orange HPS lamps of low correlated color temperature (CCT), 1900-2200 K, are being widely replaced by white LED with a higher blue content and correspondingly higher CCTs (3000-5000 K) [105, 106]. The increase in blue content of the artificial nightscape has some unintended environmental consequences. Artificial light is a relevant stimulus across all regions of the optical spectrum, see e.g. [107-109], but short wavelengths, including blue, violet and near ultraviolet, are particularly effective as visual stimuli eliciting behavioural responses across many taxa [110-113]. The current process of replacing HPS lamps by white LEDS has then the potential to produce new ecological effects even in those places where the absolute lighting levels, measured in human photopic units, are not changed.

The change from warm to white outdoor light was not motivated by any explicit demand of the lighting sector nor by any relevant visual need that could not be met with HPS lamps; it was rather a by-product of the industry attempt to keep the luminous efficacy of the phosphor-coated LED lamps comparable to the one of the HPS they intended to outphase. During the first steps of solid-state lighting deployment, harsh white LEDs were marginally more efficient than warm ones, requiring less amounts of phosphors for their fabrication, a difference that nowadays is of secondary importance. In fact the meaningful energy savings enabled by LEDs in outdoor lighting do not primarily stem from their luminous efficacy but from the fact that their light can be more precisely directed to where it is needed, decreasing the spilled light and increasing the overall utilization factor, and from their ability to regulate at will the amount of emitted light with extremely short response times. An active marketing campaign based on concepts like modernity, color rendering, scotopic vision and others, often incorrectly applied, contributed to make white light the 'new normal' in ample areas of the world.

## 5. Lighting technologies, rebound effects, and social uses of light

Many technological changes in light emitting systems are purportedly driven by the search of lamps with higher luminous efficacies, that is, higher ratios of produced visible light to consumed electrical power. It is somehow expected that this would lead, in the long term, to a reduction in the overall amount of energy used to maintain our actual lifestyles. However, the documented history of artificial lighting shows that increases of luminous efficacy tend to give rise larger amounts of emitted light (via the extension of the illuminated surface, the number of users, and the development of new uses of light), and in the long run to net increases in energy consumption [2], a rebound associated with the reduction of the effective price of light [114]. The present process of deployment of solid-state lighting provides no compelling reason for expecting a different outcome. Furthermore, LEDs are expected to have average useful lifetimes longer than gas-discharge lamps. While this is a desirable feature, it certainly puts pressure on different industrial lighting agents, since the income derived from maintenance and periodic lamp replacements is consequently reduced. This may lead to compensatory income strategies based on the extension of the illuminated surface and the intensification of uses and times of operation in those areas already saturated with streetlights. As a matter of fact there is a widespread trend of installing new lighting systems intended for ornamental, non-functional uses, including the generalized illumination of façades, monuments, bridges, beaches, cliffs, woods and even botanic gardens, not to mention the massive installation of ultra-bright LED billboards of large sizes in the central town districts or in view from high traffic roads [115]. There is also large room for increasing the urban light levels at nighttime, since the human visual system has a poor memory and can easily adapt to illuminances from a few a tenths of lx to more than $10^5$ lx. Although conjunctural shortages in energy supply and increased energy prices may slow somewhat this process during limited time periods, it is not clear whether they will effectively stop the progressive loss of the natural night. They did not stop it in previous energy crises.

The ways and extent of lighting the night are, ultimately, the result of social choices. There is no pre-defined way in which the artificial night should be lit. Our present lighting schemes are strongly tied to particular models of transportation and mobility, and specially to the double assumption that (i) vehicles should be able to run through the cities at high speed, in spaces conflicting with pedestrian mobility and (ii) increasing illumination would contribute to reduce traffic accidents. Both these assumptions are on question today, either by the emergence of new mobility and lighting paradigms [116, 117] or by the critical reappraisal of the relationship between illumination levels and safety [118-120]. As a matter of fact, basic physical principles indicate that the limitation of the total amount of emitted light is a necessary, unavoidable condition for controlling light pollution [121], whereas technical updates and correct engineering of lighting installations may be a complementary tool.

## 6. Additional remarks and conclusions

The human perception of the world is deeply rooted in visual inputs, and artificial light is not only taken for granted as a basic resource of modern societies but has a prevalent positive cultural connotation. These facts have contributed somehow to delay the correct appreciation of its actual polluting character. In this paper we argue that light pollution is a standard form of pollution and that artificial light at night should be consequently treated as a conventional environmental pollutant. The multiple unknowns that still remain about the interactions of artificial light with living matter at various size, time, and complexity scales, see e.g. [42], do not detract from the huge amount of converging evidence about the disruptive effects of anthropogenic light in the nocturnal environment of our planet. The widespread misconception that electromagnetic radiation in the optical band can only produce detrimental effects through ionizing or thermal processes overlooks the basic fact that the main interactions between optical radiation and living matter take place through photochemical processes activating visual and non-visual pathways.

In conclusion, artificial light at night is a useful resource and, at the same time, an environmental pollutant. Light pollution is caused by an increase of the concentration of light particles (photons) above the natural nighttime levels as a consequence of human activity. Light pollution is a multiscale disruptor of the nighttime environment, affecting from molecular processes to ecosystems, from the immediate vicinity of the sources up to hundreds of km away, with time scales ranging from seconds or less to decades, and with atmospheric variations superimposed on periodic daily and seasonal oscillations, within a general interannual increase trend.

**Competing Interests**

We declare no competing interests.

**Funding**

No specific funding was allocated to this work.